\documentclass[pre,preprint,superscriptaddress,showpacs,amsfonts]{revtex4}

\usepackage{graphicx}
\usepackage{dcolumn}
\usepackage{bm}

\def\be{\begin{equation}}
\def\ee{\end{equation}}
\def\bea{\begin{eqnarray}}
\def\eea{\end{eqnarray}}
\def\a{\alpha}
\def\b{\beta}

\def\L{{\cal L}}
\def\o{\omega}

\def\no{\nonumber}

\usepackage{amsmath}
\begin{document}
\title{Exact Solution of a Monomer-Dimer Problem: A Single Boundary Monomer  on a Non-Bipartite Lattice}
\author{F. Y. Wu}
\affiliation{Department of Physics, Northeastern University, Boston, Massachusetts 02115, U.S.A.}
\author{Wen-Jer Tzeng}
\affiliation{Department of Physics, Tamkang University, 251 Tamsui, Taipei, Taiwan}
\author{N. Sh. Izmailian}
\affiliation{Yerevan Physics Institute, Alikhanian Brothers 2, 375036 Yerevan, Armenia}
\date{\today}

\begin{abstract}

We solve the monomer-dimer problem on a non-bipartite lattice, the
simple quartic lattice with cylindrical boundary conditions, with
a single monomer residing on the boundary. Due to the
non-bipartite nature of the lattice, the well-known method of a
Temperley bijection of solving single-monomer problems cannot be
used. In this paper we derive the solution  by mapping the problem
onto one on close-packed dimers on a related lattice. Finite-size
analysis of the solution is carried out. We find from
asymptotic expansions of the free energy that the
central charge in the  logarithmic conformal field theory assumes the value
$c=-2$.

\end{abstract}
 \maketitle

\section{Introduction}

An outstanding unsolved problem in lattice statistics is the monomer-dimer problem.
  In this problem diatomic molecules adsorbed on a surface are
modeled as rigid dimers  occupying two adjacent sites  and
  lattice sites  not covered by dimers are
regarded as occupied by monomers. While the case of pure dimers has been solved in 1961 by Kasteleyn
\cite{Kasteleyn61} and  by Fisher and Temperley \cite{TeFi61,Fisher61},
  the general monomer-dimer problem has proven
to be computationally intractable \cite{Jerrum87}.

In 1974, Temperley \cite{Temperley74} introduced an intriguing bijection mapping
the dimer problem with a single monomer at the corner of a finite $M\times N$ lattice
to the counting problem of spanning trees on a related lattice, thereby providing an alternate
way of deducing the solution.
The method of Temperley bijection has since been extended
to the case when the monomer resides on other specific boundary sites \cite{TzWu03}.
 However, the success of the Temperley bijection apparently relies on the fact that
the lattices being bipartite; it does not work for non-bipartite  lattices.
  In this paper, we consider one  non-bipartite lattice, a rectangular lattice
with a cylindrical boundary condition.
By using an
 alternate mapping formulated recently by one of us \cite{Wu06,WuE},
we solve the monomer-dimer problem on this lattice when a single monomer resides
on the boundary.
 We also clarify the mathematical content of the solution by carrying out finite-size analysis
of the solution.

\section{Single Monomer on the Boundary of a Cylinder}

Consider a simple quartic lattice ${\cal L}$ consisting of an array of
$N$ rows and $M$ columns embedded on the surface of a cylinder with periodic boundary conditions
imposed in the horizontal direction.
 See Fig.\ 1(a) for an illustration.
For $MN$ odd, hence both $M$, $N$ odd, the lattice is not bipartite.
But the lattice can be fully covered by one monomer and $(MN-1)/2$ dimers.
 We consider the problem of evaluating its generating function when the single
monomer resides on the boundary.
\begin{figure}[ht]
\includegraphics{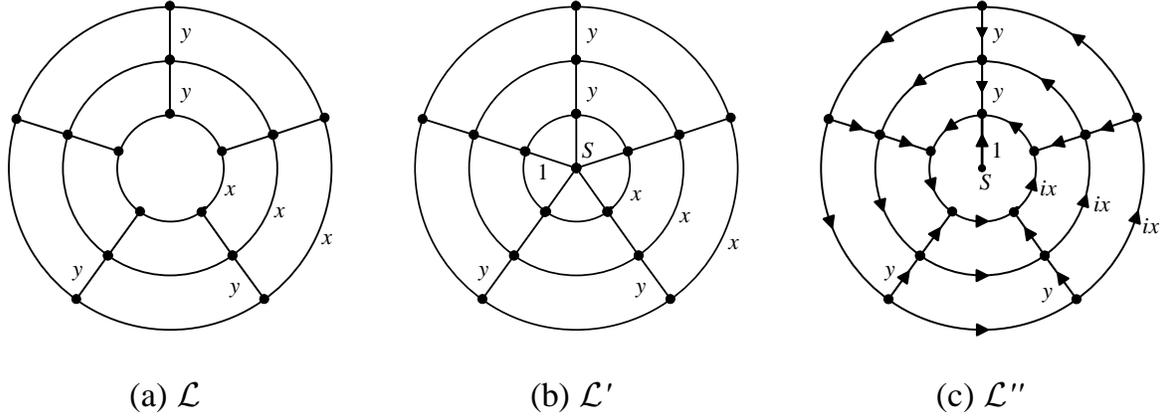}
\vskip 7.0 cm

\caption{(a) A simple quartic lattice ${\cal L}$ consisting of an array of
$N=3$ rows and $M=5$ columns embedded on the surface of a cylinder.
 (b) A self-dual lattice $\L'$ derived from $\L$ by adding a new site $S$ connected to all $M$ sites of one boundary of $\L$.
  (c) A oriented lattice $\L''$ constructed from $\L'$ by keeping only one edge connecting to $S$. A phase factor $i$
is associated to all $x$ dimers.}
\label{fig:Lattices}
\end{figure}

On first sight, one would attempt to use the  Temperley bijection of  mapping.
However, it can be readily verified that the attempt invariably fails, apparently due to the fact that  $\cal L$ is not
bipartite. Instead, we adopt an alternate formulation devised by one of us    \cite{Wu06,WuE} which does not
make use of the
Temperley bijection.

Denote the desired generating function by
\begin{equation}
G_{MD}(x,y)=\sum_{\rm config}x^{n_1}y^{n_2}, \label{gen}
\end{equation}
where the summation runs over all monomer-dimer configurations with
a single monomer on one of the two boundaries, $x>0$ and $y>0$ are the weights of, respectively, horizontal
and vertical dimers as indicated in Fig.\ 1(a), and
 $n_1$ and $n_2$  the numbers of horizontal and vertical dimers subject to $n_1 + n_2 = (MN-1)/2$.
For quick reference we first give the final result which holds for
$M,N \ge 3$, \be G_{MD}(x,y)
=2Mx^{(M-1)/2}y^{(N-1)/2}\prod_{m=1}^{\frac{M-1}{2}}\prod_{n=1}^{\frac{N-1}{2}}
\left(4x^2\sin^2\frac{2m\pi}{M}+4y^2\cos^2\frac{n\pi}{N+1}\right).\label{general}
\ee In contrast, the monomer-dimer generating function with a
single monomer on the boundary of an $M\times N$ net with free (open)
boundaries is \cite{TzWu03}
\be
G_{MD}^{\rm
free}(x,y)=(M+N-2)x^{(M-1)/2}y^{(N-1)/2} \prod_{m=1}^{\frac {M-1}
{2} }\prod_{n=1}^{\frac {N-1} {2} }
\left(4x^2\cos^2\frac{m\pi}{M+1}+4y^2\cos^2\frac{n\pi}{N+1}\right),
\label{freegen}
\ee
where the factor $M+N-2$ is the number of
equivalent boundary sites where the monomer can reside.

Results of enumerations of (\ref{general}) and (\ref{freegen}) for small lattices
are shown in Table \ref{enumeration}.

\begin{table}[htbp]
\caption{Enumerations of monomer-dimer configurations.}
\begin{tabular}{c|r|r}
    \hline
\  $M\times N$ lattice \quad &\quad $G_{MD}(1,1)$ given by (\ref{general}) \quad &
  \quad $G_{MD}^{\rm free}(1,1)$ given by (\ref{freegen}) \quad \\
    \hline
    $5\times 5$     & 3,190 \quad \quad& 1,536 \quad \quad\\
    $5\times 7$     & 53,010 \quad \quad& 24,150 \quad \quad\\
    $7\times 5$     & 56,434 \quad \quad& 24,150 \quad \quad\\
    $7\times 7$     & 3,118,178 \quad \quad& 1,204,224 \quad \quad\\
    $7\times 9$     & 171,527,426 \quad \quad& 57,961,134 \quad \quad\\
    $9\times 7$     & 165,771,810 \quad \quad& 57,961,134 \quad \quad\\
    $9\times 9$     & 29,845,632,402 \quad \quad& 8,921,088,000 \quad \quad\\
    \hline
  \end{tabular}
\label{enumeration}
\end{table}

To derive (\ref{general}) we consider first
the {\it close-packed} dimer problem on a related lattice ${\cal L}'$
constructed  from $\cal L$ by connecting all $M$
sites on one boundary to a single new site $S$ as shown in Fig.\ 1(b).
Dimers connecting boundary sites to $S$ all carry  weight $1$.
 It is of interest to note that the lattice ${\cal L}'$ is
self-dual and that the lattice has been considered previously by Lu and Wu \cite{LuWu98}
in the context of Ising partition function zeroes.

Denote the generating function of close-packed dimers on ${\cal L}'$ by $G_D({\cal L}';x,y)$.
Since in a close-packed configuration $S$ must be covered by a dimer (of weight $1$), and the
dimer must end at one of the
$M$ equivalent boundary sites which can  be regarded as being occupied by a monomer on $\L$,
there exists a correspondence between  dimer configurations on $\L'$ and
monomer-dimer configurations on $\L$. We are led to the identity
\be
G_{MD}(x,y) =  2\, G_{D}({\cal L}'; x,y), \label{1}
\ee
where the extra factor $2$ comes from the fact that there are 2 boundaries on a cylinder.

To evaluate $G_{D}({\cal L}'; x,y)$ we introduce the lattice
$\L''$ shown in Fig.\ 1(c) where $S$ is connected to only one  boundary site.
 Denote  the
generating function of close-packed dimers on $\L''$ by $G_D(\L''; x,y)$. It is clear that
we have the further identity
\be
G_D({\cal L}';x,y) = M\,G_{D}({\cal L}''; x,y). \label{2}
\ee
It remains to evaluate $G_{D}({\cal L}''; x,y)$. But this is the problem solved in \cite{Wu06,WuE}.

In the analysis given in \cite{Wu06}, close-packed dimers on a lattice similar to ${\cal L}''$ are enumerated using the
Kasteleyn approach \cite{Kasteleyn61}. Since our procedure follows closely  that discussed in
\cite{Wu06}, we give  an outline and highlight  the difference.

Orient edges of $\L''$ and associate a phase factor $i$ to all
$x$ edges as shown in Fig.\ 1(c). The  only thing new from \cite{Wu06} is that we need
to ascertain signs of all terms in the Pfaffian are the same. However, it can be shown  \cite{McWu73,note}
that this always is the case
for $M= $ odd. Then the desired generating function
$G_{D}({\cal L}''; x,y)$  is given in terms of the Pfaffian of a matrix $A'$ \cite{WuE},
\begin{equation}
i^{(M-1)/2}G_{D}(\L''; x,y) = {\rm Pf}(A') =\sqrt{\det A'}.\label{eq:Gfunc}
\end{equation}
Here $A'$ is the antisymmetric Kasteleyn matrix of dimension $(MN+1)\times(MN+1)$ for the lattice ${\cal L}''$
explicitly given by
\begin{equation}
A'=\left(\begin{array}{cc}
0 &
\begin{array}{ccccccc}
0 & \cdots  & 0 & 1 & 0 & \cdots  & 0
\end{array}
\\
\begin{array}{c}
0\\
\vdots\\
0\\
-1\\
0\\
\vdots\\
0
\end{array}
& A
\end{array}
\right),\label{Matrix:A'}
\end{equation}
where $A$ is the Kasteleyn matrix of dimension $MN\times MN$ for ${\cal L}$.
The position of the elements $\pm 1$ in the first row and column is that of the site $\{m,1\}$
connected to $S$ (see below).
Explicitly, $A$ is given by
\begin{equation}
A=ixS_M\otimes I_N+yI_M\otimes T_N, \label{ST}
\end{equation}
with $I_M$ is the $M\times M$ identity matrix, $S_M$ is the {\it periodic} $M\times M$ matrix
\begin{equation}
S_M=\left(\begin{array}{cccccc}
0&1&0&\cdots&0&-1\\
-1&0&1&\cdots&0&0\\
0&-1&0&\cdots&0&0\\
\vdots&\vdots&\vdots&\ddots&\vdots&\vdots\\
0&0&0&\cdots&0&1\\
1&0&0&\cdots&-1&0\\
\end{array}
\right),
\end{equation}
and $T_N$ is the $N\times N$ matrix
\begin{equation}
T_N=\left(\begin{array}{cccccc}
0&1&0&\cdots&0&0\\
-1&0&1&\cdots&0&0\\
0&-1&0&\cdots&0&0\\
\vdots&\vdots&\vdots&\ddots&\vdots&\vdots\\
0&0&0&\cdots&0&1\\
0&0&0&\cdots&-1&0\\
\end{array}
\right).
\end{equation}
Note that we have $T_M$ instead of $S_M$ in the corresponding expression in \cite{Wu06}.

Label elements of $A$ by $\{m,n;m',n'\}$, where $(m,n)$ specifies the column and row of the position a site.
 The determinant of the Kasteleyn matrix $A'$ can be computed by Laplace expanding  along
the first row and first column leading to
\begin{equation}
\det A'=C(A;\{m,1;m,1\}),\label{eq:detKas}
\end{equation}
where $C(A;\{m,1;m,1\})$ is the cofactor of the $\{m,1;m,1\}$ element of $A$,
and we have specified the site connecting to $S$ in Fig. 1(c) as  $\{m,1\}$.

Since the cofactor $C(A;\{m,1;m,1\})$ is proportional to the
product of the nonzero eigenvalues of the matrix $A$, we need to determine the eigenvalues of $A$.
 This is done in the next section.

\section{Eigenvalues of the Kasteleyn matrix $A$}

The matrix $S_M$ can be diagonalized by the similarity transformation
\be
V_M^{-1}S_MV_M=\Omega_M, \no
\ee
where $V_M$ and its inverse $V_M^{-1}$ are $M\times M$ unitary matrices with elements
\begin{eqnarray}
V_M(m_1,m_2)&=&\frac{1}{\sqrt{M}}e^{i2m_1m_2\pi/M},\nonumber\\
V_M^{-1}(m_1,m_2)&=&\frac{1}{\sqrt{M}}e^{-i2m_1m_2\pi/M}, \quad 1\le \{m_1,m_2\}\le M,\label{eq:Vmatrix}
\end{eqnarray}
 and $\Omega_M$ is an $M\times M$ diagonal matrix with  the eigenvalues $\omega_m$ of $S_M$ as entries,
\be
\omega_m=2i\sin\frac{2m\pi}{M}, \quad\quad 1\le m\le M\, . \no
\ee
Similarly, as  in \cite{Wu06}, the matrix $T_N$ is diagonalized by the similarity transformation
\be
U_N^{-1}T_NU_N=\Gamma_N, \no
\ee
where $U_N$ and its inverse $U_N^{-1}$ are $N\times N$ unitary matrices with elements
\begin{eqnarray}
U_N(n_1,n_2)&=&\sqrt{\frac{2}{N+1}}\,i^{n_1}\sin\bigg( \frac{n_1n_2\pi}{N+1}\bigg),\nonumber\\
U_N^{-1}(n_1,n_2)&=&\sqrt{\frac{2}{N+1}}\,i^{-n_2}\sin\bigg( \frac{n_1n_2\pi}{N+1}\bigg)\label{eq:Umatrix}
\end{eqnarray}
for $1\le \{n_1,n_2\} \le N$, and $\Gamma_N$ is an $N\times N$ diagonal matrix having eigenvalues $\gamma_n$ of $T_N$
as entries,
\be
\gamma_n=2i\cos\frac{n\pi}{N+1}, \quad\quad 1\le n\le N\,. \no
\ee
Thus, the $MN\times MN$ matrix $A$ can be diagonalized by the similarity transformation
generated by $U_{MN}=V_M\otimes U_N$, leading to
\begin{equation}
U_{MN}^{-1}AU_{MN}=\Lambda_{MN}, \label{eq:diagonalization}
\end{equation}
where $\Lambda_{MN}$ is an $MN\times MN$ diagonal matrix having eigenvalues $\lambda_{m,n}$ of $A$ as entries,
\be
\lambda_{m,n}=2i\left(ix\sin\frac{2m\pi}{M}+y\cos\frac{n\pi}{N+1}\right), \quad
1\le m\le M, \quad 1\le n\le N\, . \no
\ee
 Note that $\lambda_{m,n}$ vanishes
at $m=M, n=(N+1)/2$.
Elements of $U_{MN}$ and its inverse $U_{MN}^{-1}$ are
\begin{eqnarray}
U_{MN}\{m_1,n_1;m_2,n_2\}&=&V_M(m_1,m_2)U_N(n_1,n_2)\nonumber \\
U_{MN}^{-1}\{m_1,n_1;m_2,n_2\}&=&V_M^{-1}(m_1,m_2)U_N^{-1}(n_1,n_2). \label{U}
\end{eqnarray}

 Using the identities \ $\sin(2\pi-\theta)=-\sin\theta$ \ and
$\ \cos(\pi-\theta)=-\cos\theta$ , the product \be P\equiv
\prod_{m=1}^M \prod_{\substack{n=0\\(m,n)\not=(M,\frac {N+1} 2
)}}^{N-1} \lambda_{m,n}\,, \ee where the product excludes the zero
eigenvalue at $(m,n) =(M,\,\frac {N+1} 2)$, can be rearranged as
 \begin{equation}
P=Q\prod_{m=1}^{\frac{M-1}{2}}\prod_{n=1}^{\frac{N-1}{2}}\left(4x^2\sin^2\frac{2m\pi}{M}
+4y^2\cos^2\frac{n\pi}{N+1}\right)^2, \label{eq:DefP}
\end{equation}
where the factor $Q$ collects all factors with either  $n=(N+1)/2$ or $m=M$,
namely,
\begin{eqnarray}
Q&=&\prod_{m=1}^{\frac{M-1}{2}}\left(-4x^2\sin^2\frac{2m\pi}{M}\right)
\prod_{n=1}^{\frac{N-1}{2}}\left(4y^2\cos^2\frac{n\pi}{N+1}\right) \nonumber \\
&=&(-1)^{(M-1)/2}\bigg( \frac{M(N+1)}{2}\bigg)x^{M-1}y^{N-1}, \label{eq:DefQ}
\end{eqnarray}
after  using  the identities
\be
\prod_{m=1}^{\frac{M-1}{2}}\left(4\sin^2\frac{2m\pi}{M}\right)=M,\quad
\prod_{n=1}^{\frac{N-1}{2}}\left(4\cos^2\frac{n\pi}{N+1}\right)=\frac{N+1}{2},\quad \mbox{$M,N $ odd}. \no
\ee
The expressions (\ref{eq:DefP}) and (\ref{eq:DefQ}) apply to $M,N\ge 3$ and will be used in the next section.

\section{Evaluation of The Generating Function (\ref{gen})}

We now compute the generating function (\ref{gen}).

Combining (\ref{1})-(\ref{eq:Gfunc}) with (\ref{eq:detKas}), we obtain the following expression,
\be
G_{MD}(x,y) = 2 \,M \,i^{(1-M)/2} \sqrt { C(A;\{m,1;m,1\}) }\, .
\ee
where $ C(A;\{m,1;m,1\}) $ is the cofactor of the $(m,1; m,1)$ element of the matrix $A$.

The computation of cofactors of a singular matrix like $A$ requires special attention since the matrix
does not possess an inverse.  The difficulty was resolved   in
\cite{Wu06}
  by  perturbing the matrix $A$ slightly rendering it non-singular to permit an inverse.
 By carrying out this analysis details of which we refer to \cite{Wu06}, one finds
the   cofactor
\be
C(A;\{m,n;m',n'\}) = \bigg[U_{MN}\big(m',n'; M, \frac {N+1} {2}\big)
U^{-1}_{MN} \big(M, \frac{N+1} 2; m, n\big)\bigg]\, P \, ,\label{uv}
\ee
where $U_{MN}$ is the matrix   diagonalizing $A$.
Note that the index $\{ M, \frac {N+1} 2\}$ is that of the zero eigenvalue.

Elements of $U_{MN}$ and $U^{-1}_{MN}$ are given in  (\ref{U}). After combining with
(\ref{eq:Vmatrix}) and (\ref{eq:Umatrix}), we obtain from (\ref{uv})
\begin{equation}
C(A;\{m,n;m',n'\})=\bigg[  \frac{2\,i^{n'-n}} {M(N+1)} \sin\frac{n\pi}{2}\sin\frac{n'\pi}{2}\bigg]\, P
\label{eq:CofactorValue}
\end{equation}
valid for general $m,n,m',n'$.

Finally, we combine (\ref{1})-(\ref{eq:Gfunc}) with (\ref{eq:detKas}) and
(\ref{eq:CofactorValue}) at  $\{m' = m, n'=n = 1\}$,
and arrive at the expression
\be
G_{MD}(x,y)
=2\,M\,i^{(1-M)/2}\sqrt{\frac{2P}{M(N+1)}}.
\ee
This yields the generating function  (\ref{general}) given in Sec. II after
substituting with $P$ given by (\ref{eq:DefP}) and $Q$  by  (\ref{eq:DefQ}).
We note  that the result is independent of $m$ as it should.

Then, with the help of the relations
\be
\prod_{n=1}^{\frac{N-1}{2}}F\left(\cos^2\frac{n\pi}{N+1}\right)
=\prod_{n=1}^{\frac{N-1}{2}}F\left(\sin^2\frac{n\pi}{N+1}\right)
\quad \mbox{and} \quad
\prod_{m=1}^{\frac{M-1}{2}}F\left(\sin^2\frac{2m\pi}{M}\right)
=\prod_{m=1}^{\frac{M-1}{2}}F\left(\sin^2\frac{m\pi}{M}\right), \no
\ee
valid for any function $F(\cdot)$,
the generating function (\ref{general}) can be written in the
equivalent form,
\be
 G_{MD}(x,y) =2M x^{(M-1)/2}
y^{(N-1)/2}\prod_{m=1}^{\frac{M-1}{2}}\prod_{n=1}^{\frac{N-1}{2}}
\left(4x^2\sin^2\frac{m\pi}{M}+4y^2\sin^2\frac{n\pi}{N+1}\right),
\quad M, N = {\rm odd}.\label{general1}
\ee

It is convenient at this point to introduce a function
\bea
H(z; M, N) &\equiv & \bigg[\prod_{m=0}^{M-1}
\prod_{\substack{n=0\\(m,n)\not=(0,0)}}^{N-1}
\left(4z^2\sin^2\frac{m\pi}{M}+4\sin^2\frac{n\pi}{N}\right)
\bigg]^{1/2} \no \\
&=& z^{MN-1} H( 1/z; N,M), \quad\quad {\rm any\>\>} M,N >1. \label{EE1}
\eea

It will be shown in Appendix A that  we have
 \be
G_{MD}(x,y) =
R_{M,N}(y,z)\, \sqrt {H(z; M, N+1)},\quad \quad M, N = {\rm odd},
\label{Gcyl}
 \ee
 where $z= x/ y$ and
\bea
[R_{M,N}(y,z)]^{\,2} &=&
\frac {4M y^{M N-1}} {(N+1)z^M\,S_M(z)}, \no\\
 S_M(z)&=&\sinh\big(M {\sinh}^{-1} (1/z) \big).\no
 \eea
The advantage of using (\ref{Gcyl}) instead of (\ref{general1}) for the
generating function is that the factor $R_{M,N}(y,z)$
sorts out major contributions in the asymptotic expansions of the free energy (\ref{FM})
and (\ref{FN}) discussed below.

Two equivalent expressions of $H(z; M, N+1)$ can be obtained by
taking one of the products in (\ref{EE1}) in a closed form. Taking
the product over $n$, we obtain
\begin{equation}
H(z;M,N+1)=(N+1) \prod_{m=1}^{M-1}2 \textstyle{~\!{\rm
sinh}\left[(N+1)\omega_z\!\left(\frac{m\pi}{M}\right)\right] }
\label{H1}
\end{equation}
where \be \o_z(k)= \sinh^{-1} (z\sin k) \label{dispersion} \ee is
the lattice dispersion relation, and  we have used the identities
(\ref{C1}) and (\ref{C3}).

 Similarly, taking the product over $m$  and making use of (\ref{C3}) and the equivalence
(\ref{EE1}),  we obtain
\begin{equation}
H(z;M,N+1)=M z^{M(N+1)-1}\,\prod_{n=1}^{N} 2 \textstyle{~\!{\rm
sinh}\left[M\omega_{1/z}\!\left(\frac{n \pi}{N+1}\right)\right] }. \label{H2}
\end{equation}

\section{Finite-Size Analysis and Asymptotic Expansions}

Define the ``free energy" of the monomer-dimer system as
\bea
F_{M,N}(x,y) &=&  -\ln G_{MD}(x,y) \no \\
&=&
- \ln R_{M,N}(y,z)  -\frac 1 2 \ln H (z; M, N+1),\label{FMN}
\eea
where we have made use of (\ref{EE1}).  We note that
other than an overall factor\,   $\big[4\sin^2(\a\pi/M) +
4\sin^2(\b\pi/N)\big]$,
the function $H(z;M,N+1)$ is the special
case of $\a = \b = 0$ of a more generally defined function $ Z_{\alpha,\beta}(z; M, N+1)$
 introduced, and  analyzed  in details in \cite{Izma03,Ivash02}.
This permits us to use   results of  \cite{Izma03,Ivash02} to  write down a general
expression for $F_{M,N}(x,y)$,
which we shall not reproduce.
Instead, we focus  on the free energies
\be
 F_M =  \lim_{N\to \infty} \frac 1 N F_{M, N}(x,y) \quad
{\rm and}\quad F_N =  \lim_{M\to \infty} \frac 1 M F_{M, N}(x,y) \no
\ee
of infinite ``strips" and their asymptotic expansions.

The asymptotic expansions can
be deduced by applying the Euler-MacLaurin summation identity to $\ln H(z;M,N+1)$.
Using $H(z;M.N+1)$ given by (\ref{H1}) and (\ref{H2}), respectively, we obtain from (\ref{FMN})
using (\ref{H1}) and (\ref{H2}), respectively,
\bea
F_M
&=&  -\frac{M}{2}\ln y- \frac{1}{2}\sum_{m=1}^{M-1}
\textstyle{~\!\omega_z\! \left(\frac{m\pi}{M}\right)} \no \\
&=&M f_{\rm bulk} +\sum_{p=1}^\infty\left(\frac{\pi}{
M}\right)^{2p-1}\frac{d_{2p-2}(z)}{(2p-2)!}\bigg(\frac{B_{2p}}{2p}\bigg)\no\\
 &=&M f_{\rm
bulk}+\frac{\pi z}{12}\bigg(\frac{1}{M}\bigg)+\cdots, \quad ({\rm infinite \>\> length}),\label{FM}
\eea
\bea
F_N &=&
-\frac{N}{2} \ln (yz)  + \frac  1 2 \sinh^{-1} (1/z) -\frac{1}{2}\sum_{n=1}^{N}
\textstyle{~\!\omega_{1/z}\! \left(\frac{n\pi}{N+1}\right)} \no \\
  &=& N f_{\rm bulk}+2 f_{\rm surface}
+\sum_{p=1}^\infty\left(\frac{\pi}{ N+1}\right)^{2p-1}\frac{
d_{2p-2}(1/z)}{(2p-2)!}\bigg(\frac{B_{2p}}{2p}\bigg)
\nonumber \\
 &=&N f_{\rm bulk}+2f_{\rm surface}+\frac{\pi}{12 z}\bigg(\frac{1}{
N+1}\bigg)-\cdots, \quad ({\rm infinite \>\> perimenter}). \label{FN}
\eea
where
\bea
f_{\rm bulk} &=& -\frac {1}{2}\ln y - \frac  {1}{2\pi}\int_0^\pi \omega_z(k)dk \no \\
      &=& -\frac {1}{2}\ln( yz) - \frac  {1}{2\pi}\int_0^\pi \omega_{1/z}(k)dk ,\no \\
  f_{\rm surface}&=&\frac {1}{4 } \sinh^{-1} (1/z)  - \frac  {1}{4\pi}\int_0^\pi \omega_{1/z}(k) dk, \no
\eea
$d_{2p}(z) $ are the coefficients in the Taylor expansion
\be
\omega_z(k)=\sum_{p=0}^\infty\frac{d_{2p}(z)}{(2p)!}k^{2p+1}, \no
\ee
with\, $d_0(z)=z$,\,  $d_2(z) =-z(1+z^2)/3$,\, $d_4(z)=z(1+z^2)(1+9z^2)/5, \cdots,$
and\, $B_2 = 1/6,\, B_4 = -1/30,\, B_6= 1/42, \cdots, $ are the Bernoulli numbers.
The two equivalent expressions of $f_{\rm bulk}$ are obtained from (\ref{FM}) and (\ref{FN}), respectively.
We remark that the equivalence of the two expressions is verified by the intriguing integral identity
\be
\frac  {1}{\pi}\int_0^\pi \big[ \textstyle{~\!\sinh^{-1}\, \! (z\sin\theta )}
- \textstyle{~\!\sinh^{-1} \! \left(\frac{1}{z}\sin \theta \right)} \big] \,  d\theta = \ln z  \label{ID}
\ee
obtained by noting that the derivative of the left-hand side of (\ref{ID}) with respect to $z$ reduces to  $1/z$
after carrying out the integration.

The general theory of finite-size analysis \cite{blote,aff,cardy}
dictates that  the free energy per unit length of a lattice model
at criticality on an infinitely long strip of width $\cal N$ assumes
the form \cite{cardy}
\begin{equation}
F_{\cal N} ={\cal N} f_{\rm bulk}  +  f_{\rm surface} + \frac{\Delta}{\cal N} + \cdots,
\label{free}
\end{equation}
 in an asymptotic expansion where $f_{\rm bulk}$ and $f_{\rm surface}$ are free energy densities of the order of O(1)
and $\Delta$ is a constant. Unlike the free energy densities, the constant
$\Delta$ is universal and its value is related to the
central charge \,$c$ \,in the logarithmic conformal field theory in a relation which
depends on the boundary
conditions in the transversal direction.  Explicitly,
$\Delta$ is proportional to an effective central charge $c_{\rm eff} =
c - 24\,h_{\rm min}$, where\,  $c$ \,is the central charge  characterizing
the universality class of the lattice model,  as \cite{blote,IzmPrRuHu1}
\bea
 \Delta &=& -{\pi\zeta \over
6}\, c_{\rm eff} =- {\pi\zeta \over
6}\big(c - 24\, h_{\rm min}\big)
\;\; \mbox{on a cylinder of infinite length}, \label{Astrip} \\
 \Delta &=& - {\pi \zeta\over {24}}\, c_{\rm eff}  = - {\pi \zeta\over {24}}  \big( c - 24\, h_{\rm min} \big)
\;\; \mbox{on a cylinder of infinite perimeter}, \label{Acyl}
\eea
where the number $h_{\rm min}$ is the smallest conformal weight in the
spectrum of the Hamiltonian with the given boundary conditions and
$\zeta$ is an anisotropy factor. In our case we find
from (\ref{FM}) and (\ref{FN}) that
$\zeta = z$ and $1/z$, and
$\Delta = \pi z/12$\,  and\, $ \pi /12z$, respectively,
in  (\ref{Astrip}) and (\ref{Acyl}).

To retain the characteristics of a monomer on the surface, we
consider  a cylinder of infinite perimeter in a geometry
which retains two surfaces. Therefore
we use (\ref{Acyl}) and (\ref{FN}), or $F_N$,  for which the boundary condition in the
transverse direction is free (open) boundaries.  It is known \cite{ruelle}
that for free (open) boundaries $h_{\rm min} = 0$. Hence
we deduce the central charges
\be
c=c_{\rm eff} = -2.  \label{centralcharge}
\ee

On the other hand, if one uses
(\ref{FM}), or $F_M$, the system is an infinitely long cylinder  with
a perimeter $M$. The two physical boundaries of the lattice are located at
infinity so the existence of a monomer on the boundary is immaterial.
The situation reduces  to that of a pure dimer problem studied in \cite{IzmPrRuHu1}.
For $M=$ odd  we are considering, the analysis of \cite{IzmPrRuHu1} also gives\,
$\Delta = \pi \zeta / 12$\, as in (\ref{FN}).
However, for $M$ odd, the  boundary in the transverse direction is ``frustrated" requiring
special attention.
It is argued in \cite{IzmPrRuHu1}
that in this case one should  use (\ref{Acyl}) with $h_{\rm min} = 0$.
This again leads  to the same central charges (\ref{centralcharge}).

 We remark that the\, $c=-2$ \,central
charge has been reported previously \cite{TzWu03} in the solution (\ref{freegen})
of a single monomer on the surface of a rectangular net with free (open)
boundaries.

\section{Summary}
We have derived the closed-form expression of the monomer-dimer
generating function for a non-bipartite  rectangular lattice under cylindrical
boundary conditions with a single monomer confined to reside on the boundary. We have also carried out
a  finite-size analysis of the free energy.  Asymptotic expansions of the free energy of
strips  of infinite lengths in
 the periodic and free (open) directions are obtained using the Euler-MacLaurin summation
formula. We find the central charge in the framework of the logarithmic
conformal field theory to be\, $c = -2$.

\section{Acknowledgments}
We are grateful to Prof. H. W. J. Bl\"ote for insightful comments on the role of frustrated
boundaries in finite-size analysis.
The work of WJT is supported in part by the National Science
Council of the Republic of China under Grant No. NSC
97-2112-M-032-002-MY3.
The work of NSI is supported in part by
National Center for Theoretical Sciences: Physics Division,
National Taiwan University, Taipei, Taiwan.
We thank Dr. Maw-Kuen Wu for hospitality at the Institute of Physics, Academia Sinica, Taipei,
where this work was initiated and completed.

\appendix
\section{}
\label{A}
In this appendix we establish the expression (\ref{Gcyl}) for the
generating function.

First, we rewrite the generating function (\ref{general1}) as
 \be
G_{MD}(x,y)=2 M z^{\frac{M-1}{2}}y^{\frac{M
N-1}{2}}\prod_{m=1}^{\frac{M-1}{2}}\prod_{n=1}^{\frac{N-1}{2}}g(m,n), \label{eq:G}
\ee
where
\[
g(m,n)\equiv
4\left(z^2\sin^2\frac{m\pi}{M}+\sin^2\frac{n\pi}{N+1}\right).
\]
To  extend the limits of the
products in (\ref{eq:G}) to $M-1$ and $N$ as in (\ref{EE1}), we note
\begin{eqnarray*}
{\prod_{m=0}^{M-1}\prod_{\substack{n=0\\(m,n)\not=(0,0)}} ^{N}}g(m,n)=\,C_1\, C_2, C_3
\left[\prod_{m=1}^{\frac{M-1}{2}}\prod_{n=1}^{\frac{N-1}{2}}g(m,n)\right]^4\,\quad M, N = {\rm odd} ,
 \end{eqnarray*}
where $C_1, C_2, C_3$ collect respective products for $m=0$, $n=0$, and $\{m\not=0, n=(N+1)/2\}$. Namely,
for $M,N$ odd,
\bea
C_1 &=& \prod_{n=1}^N g(0,n) = \prod_{n=1}^N \bigg( 4 \sin^2 \frac {n\pi} {N+1} \bigg)
  = (N+1)^2 ,\quad N\ge 1, \label{C1}\\
C_2 &=&  \prod_{m=1}^{M-1}g(m,0)=\prod_{m=1}^{M-1} \bigg(4z^2\sin^2\frac {m\pi}{M}\bigg)=  M^2z^{2(M-1)},
   \quad M>1   , \label{C2} \\
C_3 &=&  \prod_{m=1}^{M-1}g\bigg(m,\frac{N+1}{2}\bigg) = z^{2M} \sinh^2\bigg[M\sinh^{-1} \bigg(\frac 1 z\bigg)\bigg],
\quad M>1 ,
 \label{C3}
\eea
where the product (\ref{C3}) is a special cases of the
identity \cite{Grad94}
\be
\prod_{m=0}^{M-1}\left(4\sinh^2\theta+4\sin^2\frac{m\pi}{M}\right)=4\sinh^2(M\theta), \quad M\ge 1.
\label{identity}
\ee
Combining these results, the generating
function (\ref{eq:G}) reduces to (\ref{Gcyl}).


\begin{thebibliography}{18}
\bibitem{note} Technically, this is because no transposition polygon obtained by superimposing two dimer
configurations, which is always of even length, can loop around the cylinder for $M=$ odd. It follows that we do not need
to worry about the periodic boundary condition, and the signs of all terms in the Pfaffian are  the same as
determined in  \cite{Wu06}. Furthermore, since the monomer resides on the boundary, we do not need to worry
about
transposition polygons  looping around the monomer which, if existing, change the sign of some terms.
\end{thebibliography}
 \end{document}